\documentstyle[12pt]{article} 
\input epsf
\def\baselinestretch{1.3}
\newcommand{\ba}{\begin{array}}
\newcommand{\ea}{\end{array}}
\newcommand{\bd}{\begin{displaymath}}
\newcommand{\ed}{\end{displaymath}}
\newcommand{\be}{\begin{equation}}
\newcommand{\ee}{\end{equation}}
\newcommand{\bea}{\begin{eqnarray}}
\newcommand{\eea}{\end{eqnarray}}
\newcommand{\KR}{Kalb-Ramond} 

\def\a{\alpha}

\def\l{\lambda}
\def\m{\mu}
\def\n{\nu}

\def\s{\sigma}
\def\p{\pi}

\def\q2 {q^2}

\begin{document}
\begin{flushright}
hep-th/0403098
\end{flushright}

\begin{center}
{\large \bf  Bulk Kalb-Ramond field in Randall Sundrum scenario}\\[15mm]
Biswarup Mukhopadhyaya\footnote{E-mail: biswarup@mri.ernet.in}\\
{\em Harish-Chandra Research Institute,\\
Chhatnag Road, Jhusi, Allahabad - 211 019, India}\\

Siddhartha Sen \footnote{E-mail: sen@maths.tcd.ie} \\
{\em School of Mathematics, Trinity College\\ 
University of Dublin, Dublin 2, Ireland}\\

Somasri Sen \footnote{E-mail: somasri@cosmo.fis.fc.ul.pt} \\
{\em CAAUL, Departamento de Fisica da FCUL,\\
 Campo Grande, 1749-016 Lisboa, Portugal}\\

Soumitra SenGupta \footnote{E-mail: tpssg@iacs.res.in} \\
{\em Department of Theoretical Physics\\
Indian Association for the Cultivation of Science\\
Kolkata - 700 032, India}\\[15mm] 
\end{center}
\begin{abstract}
We have considered the most general gauge invariant five-dimensional action 
of a second rank antisymmetric Kalb-Ramond tensor gauge theory, including
a topological term of the form $\epsilon^{ABLMN}B_{AB}H_{LMN}$ 
in a Randall-Sundrum scenario. Such a tensor field $B_{AB}$ (whose rank-3
field strength tensor is $H_{LMN}$), which appears in the 
massless sector of a heterotic string theory, is assumed to coexist with 
the gravity in the bulk. 
The third rank field strength corresponding to the Kalb-Ramond field   
has a well-known geometric interpretation as the spacetime torsion. 
The only non-trivial classical solutions corresponding to the effective 
four-dimensional action are found to be self-dual or anti-selfdual Kalb-Ramond 
fields. This ensures that the four-dimensional effective action on the brane 
is parity-conserving. The massive modes for both cases, lying in 
the TeV range, are related to the fundamental parameters of the theory. 
These modes can be within the kinematic reach of forthcoming TeV scale 
experiments. However, the couplings of the massless as well as massive 
Kalb-Ramond modes with matter on the visible brane are found to be 
suppressed vis-a-vis that of the 
graviton by the warp factor, whence the conclusion is that 
both the massless and the massive torsion modes  appear much 
weaker than  curvature to an observer on the visible brane.     

\end{abstract}

\vskip 1 true cm

\newpage
\setcounter{footnote}{0}

\def\baselinestretch{1.8}

An interesting idea floated in recent times is that our observable 
universe lies on a (3+1)-dimensional brane embedded in a 
higher dimensional ('bulk') spacetime. Such theories,
involving  compact spacelike extra dimensions, have shot into prominence 
by providing a natural explanation of the Planck-weak hierarchy problem. 
The possible presence of TeV scale effects in such models have made this study
worthwhile.

Theories of this type fall into two broad categories. One category 
includes the kind of models 
proposed by Arkani-Hamed, Dimopoulos and Dvali\cite{add} where a
factorisable geometry of the compact dimensions is postulated, and the 
Planck scale gets related to the TeV-scale `bulk' gravitational coupling 
via the volume of the internal dimensions. In the other category we have the
 proposal by Randall and Sundrum\cite{rs}. 
Here the hierarchy is generated by an exponential function,
called the warp factor, introduced in the metric in a non-factorisable
geometry. In minimal versions of both the scenarios,
all matter fields reside on the branes while gravity propagates in the bulk. 
The observable effects are generated by the Kaluza-Klein 
projection of the bulk gravity on the brane, where the spacings of the 
modes and the interactions with the standard model (SM) fields 
are governed by the dynamics of the respective models.

The present work concerns  the Randall-Sundrum scenario. Here, 
the extra dimension is a $Z_2$ orbifold of radius $r_c$.  There are
two branes at the two orbifold fixed points, viz. 
$\phi~=~0$ and $\phi~=~\pi$, where the compact 
co-ordinate $x_4$ is given by $x_4 ~=~ \phi r_c$. All visible matter
resides on the brane at  $\phi~=~\pi$ (called the `visible brane'or 
'TeV brane'), while gravity peaks on the brane at $\phi~=~0$ (called the 
'Planck brane' or the 'hidden brane'). The five-dimensional 
metric in this scenario can be written as 

\begin{equation}
ds^2=e^{-2\sigma}\eta_{\mu\nu}dx^{\mu}dx^{\nu}+
r_c^2d\phi^2,~~~-\pi\leq\phi\leq\pi
\end{equation}

\noindent
with $\eta_{\mu\nu}~=~(-,+,+,+)$, $\sigma~=~k{r_c}|\phi|$ and $k$ is 
of the order of the Planck mass $M_P$. The Planck scale in four 
dimension $(M_P)$ and higher dimension $(M)$ is related through      

\begin{equation}
M_P^2=\frac{M^3}{k}[1-e^{-2kr_c\pi}]
\end{equation}

\noindent
$M_P$, $M$ and $k$ are all of the same order of magnitude. In this 
warped spacetime five-dimensional mass scales are related to the 
effective four-dimensional masses 
through the `warp factor' $e^{-kr_{_c}|\phi|}$. 
For $k r_c~\simeq~12$, TeV scale mass parameters of the form 
$m~=~Me^{-kr_{_c}\pi}$ thus arise from the Planck scale on the 
`standard model' brane ($\phi=\pi$).

Implications of other types of bulk fields, such as scalars, gauge 
fields and fermions, \cite{gw}-\cite{addf} have already been explored 
as  an extension of the above theory. We focus in this paper on the case
where bulk spacetime is attributed with torsion, in the same way as in
Einstein-Cartan theories of four-dimensional gravity. 
The consequences of such bulk torsions 
have earlier been studied from different angles 
\cite{leb1,leb2,prd,prl}. In the last of these references, we studied the 
consequence of relating torsion to the rank-2 antisymmetric 
Kalb-Ramond field that occurs as a massless excitation in string theories.
Here we carry that study to a conclusion, using the most general
five-dimensional action as our starting point, and coming up with
some novel observations regarding the KR field itself in such scenarios.

As torsion 
is an inescapable consequence of matter field with spin, studies 
preoccupied with extra dimensions have   sometimes related it
\cite{leb1,leb2} to fermion fields residing 
either on the brane or in the bulk. In a different approach\cite{prd,prl} 
motivated by string theories \cite{gsw}, torsion is considered to have the 
same status as gravity in the bulk.  
Such consideration can be justified by remembering that
torsion is commonly incorporated into the theory by adding 
an antisymmetric part in the affine connection and can couple to 
all matter fields with spin. This is the essence of Einstein Cartan 
type theories\cite{hehl}. In our approach, the source of torsion
is traced to the rank-2 antisymmetric 
Kalb Ramond (KR) field $B_{MN}$, which arises as a massless mode 
in heterotic string theories. Thus we require the graviton and the
KR field to coexist in the bulk. In a sense, this amounts to 
consigning to the bulk both the characteristics of spacetime, namely, 
curvature and torsion.

To be precise, the rank-3 antisymmetric tensor required for
introducing torsion is found in $H_{MNL}$ which  is 
taken to be the strength of the KR field $B_{MN}$\cite{kr}:

\begin{equation} 
 H_{MNL} ~=~ \partial_{[M} B_{NL]}
\end{equation} 

\noindent
where  both $H_{MNL}$ and  $B_{MN}$ lie in the bulk while
matter fields reside on brane. In this approach,  
we have analysed earlier both the 
ADD\cite{prd} and RS\cite{prl} scenarios. A striking consequence of 
considering bulk torsion in RS scenario is that the torsion zero mode 
is enormously suppressed through the warp factor on the visible brane 
with respect to gravity and thus could provide
the illusion of the torsionless universe.

In our earlier analysis of the RS scenario, we have considered only the 
minimal  way of including torsion in the 5 dimensional action:

\begin{equation}
{\cal S}_G=\int d^4x\int d\phi~\sqrt{-G}~2[M^3~ R(G)-H_{MNL}~H^{MNL}]
\end{equation}

In this work we would like to extend the 5 dimensional action with a term 
of the form $\epsilon^{ABMNL} B_{AB}H_{MNL}$ so that we get the most general 
5 dimensional action which is invariant under the Kalb-Ramond gauge 
transformation, namely, ($\delta B_{MN}=\partial_{[M}\lambda_{N]}$)
\cite{pmssg}:     
\begin{equation}
{\cal S}_G=\int d^4x\int d\phi~\sqrt{-G}~2[M^3~ R(G)-H_{MNL}~H^{MNL}-
2M_0\epsilon^{ABMNL} B_{AB}H_{MNL}]
\end{equation}
\noindent
Here $M_0$ is a coupling parameter with the 
dimension of mass ($\sim$ five dimensional Planck mass $M$). At this 
point it is worth mentioning that such a term is purely 
topological in five dimensional KR gauge sector and is analogous to the Chern-Simons term $\epsilon^{ijk}A_iF_{jk}$ in 
the $U(1)$ gauge sector in three dimension. We thus explore in this work 
the effect of such a topological term on the effective four dimensional 
compactified theory in a 
Randall-Sundrum compactification scenario. One also finds in the literature 
similar-looking terms constructed out of a $U(1)$ gauge field in 
N=2, D=5 Supergravity theories \cite{sugra}. Here we focus  
on such a term originating in the Kalb-Ramond gauge sector.
Another interesting 
point to note here is that a term of the above appearance in N=2, D=5
Supergravity theories 
would break parity in even space-time dimensions, and thus 
it may be curious to check whether the inclusion of KR topological term
also leads to violation of parity when dimensional compactification 
is performed.  
If it does, then it would lead to a fresh source of parity violation
in the domain of gravity. It has been shown that such a source of 
parity violation may lead to phenomena like fermion helicity flip 
and Cosmic Microwave background anisotropy\cite{souanin}.

We start with the KR part  of the above 
five-dimensional action which, upto a dimensionless multiplicative factor,
can be written as
\begin{equation}
{\cal S}_H=\int d^4x\int d\phi~\sqrt{-G}~[H_{MNL}~H^{MNL}+
2M_0\epsilon^{ABMNL} B_{AB}H_{MNL}]
\end{equation}
\noindent
where $H_{MNL}$ is related to $B_{MN}$ by equation (3). We follow 
the convention where Latin indices run from 0 to 4 whereas Greek 
indices run from 0 to 3. Furthermore we use the KR gauge freedom to
set $B_{4\mu}=~0$. This allows us to get rid of massive vector modes on 
the brane. Thus we are left with the KR components corresponding to brane 
indices only, which, however, are functions of both compact and 
noncompact coordinates. 

Using the metric given in (1) we write the above 
action in the following form
\bea
{\cal S}_H=\int d^4x\int d\phi~r_c~e^{2\sigma(\phi)}&[&\eta^{\mu\alpha}
\eta^{\nu\beta}
\eta^{\lambda\gamma}H_{\mu\nu\lambda}H_{\alpha\beta\gamma}-
\frac{3}{r_c^2}e^{-2\sigma(\phi)} \eta^{\mu\alpha}\eta^{\nu\beta}
B_{\mu\nu}\partial_\phi^2
~B_{\alpha\beta}\nonumber\\
&+&\frac{6M_0}{r_c}~e^{-2\sigma(\phi)}~{\cal{E}}^{5\mu\nu\alpha\beta}~
B_{\alpha\beta}\partial_\phi B_{\mu\nu}]
\eea 

\noindent
Here we have used the relation 
\be
\epsilon^{MNLAB}=\frac{{\cal{E}}^{MNLAB}}{\sqrt{-G}}
\ee
\noindent
where ${\cal{E}}^{MNLAB}$ is the Levi-Civita tensor density in 
5 dimension while $\epsilon^{MNLAB}$ is the tensor formed from the 
Levi-Civita tensor density\cite{wein}. 

Assuming the following decomposition for  the \KR ~field\cite{prl}:
 
\begin{equation}
B_{\mu\nu}(x,\phi)=\sum_{n=0}^{\infty}~B^n_{\mu\nu}(x)\frac{\chi^n(\phi)}
{\sqrt{r_c}}
\end{equation}
\noindent
one can recast the five dimensional action in (7) as
\bea
{\cal S}_H=\int d^4x~\sum_{n=0}^{\infty}~
[~\eta^{\mu\alpha}\eta^{\nu\beta}
\eta^{\lambda\gamma}H^n_{\mu\nu\lambda}H^n_{\alpha\beta\gamma}&+
&3\eta^{\mu\alpha}\eta^{\nu\beta}B^n_{\mu\nu}B^n_{\alpha\beta}
(-\frac{e^{-2\sigma}}{r_c^2}~\frac{1}{\chi^n}\partial_\phi\partial_
\phi\chi^n)\nonumber\\
&+&6{\epsilon}^{\mu\nu\alpha\beta}B^n_{\mu\nu}B^n_{\alpha\beta}
(\frac{e^{-2\sigma}}{r_c}~\frac{M_0}{\chi^n}\partial_\phi\chi^n) ]
\eea
\noindent 
provided that the modes $\chi^n$ satisfy the orthonormality condition

\begin{equation}
\int_{-\pi}^\pi e^{2\sigma(\phi)} \chi^m(\phi)\chi^n(\phi)d\phi=\delta_{mn}
\end{equation}

In equation (10) with a 4D Minkowski metric, the Levi-Civita tensor 
density is replaced by the 4 dimensional Levi-Civita tensor with the help 
of the relation mentioned in (8). It is interesting to notice that the 
effective action in four-dimensions contains, apart from the kinetic term 
and the mass term ($B^n_{\mu\nu}B^{n~\mu\nu}$) for the \KR ~field, an 
additional term of the form $B^n_{\mu\nu}\tilde{B}^{n~\mu\nu}$ where 
$\tilde{B}^{n~\mu\nu}~(=\epsilon^{\mu\nu\alpha\beta}B^n_{\alpha\beta})$ 
is the dual of \KR ~field. 

On solving the equation of motion from this effective four dimensional 
action (10), it is quite straightforward to find the solution for the 
\KR ~field. We found that the only non-trivial solution corresponds to 
self-dual or anti-dual \KR ~fields i.e, 
$B^n_{\mu\nu}=\tilde{B}_n^{\mu\nu}$ or $B^n_{\mu\nu}=
-\tilde{B}_n^{\mu\nu}$. We have presented the proof of 
this result in the appendix. Such self-dual or anti self-dual conditions
imply the reduction in the degrees of freedom of the KR field.
Such a reduction originates in the additional five dimensional
topological term and will have crucial consequences, 
as will be demonstrated in what follows.

The above conclusion tells us that, in terms of the classical solutions for 
the KR tower $B^n_{\mu\nu}$, the effective four dimensional action (10) 
can be finally expressed as
\begin{equation}
{\cal S}_H=\int d^4x~\sum_{n=0}^{\infty}~
[\eta^{\mu\alpha}\eta^{\nu\beta}
\eta^{\lambda\gamma}H^n_{\mu\nu\lambda}H^n_{\alpha\beta\gamma}+
3m_{n\pm}^2\eta^{\mu\alpha}\eta^{\nu\beta}B^n_{\mu\nu}B^n_{\alpha\beta}]
\end{equation}
\noindent
where $m_{n\pm}^2$ satisfies the following differential equations,
\begin{equation}
-\frac{1}{r_c^2}\frac{d^2\chi^n}{d\phi^2}+
\frac{2M_0}{r_c}\frac{d\chi^n}{d\phi}=m_{n+}^2\chi^n e^{2\sigma}
\end{equation} 
\noindent
for the dual \KR ~field ($B^n_{\mu\nu}=\tilde{B}_n^{\mu\nu}$) and 
\begin{equation}
-\frac{1}{r_c^2}\frac{d^2\chi^n}{d\phi^2}- 
\frac{2M_0}{r_c}\frac{d\chi^n}{d\phi}=m_{n-}^2\chi^n e^{2\sigma}
\end{equation} 
\noindent  
for the antidual \KR field ($B^n_{\mu\nu}=-\tilde{B}_n^{\mu\nu}$). Evidently,
$\sqrt{3}m_{n+}$ and $\sqrt{3}m_{n-}$ gives the mass of the nth mode of 
the self-dual field and anti-dual field respectively. An interesting point 
to be noted here is that in the effective four-dimensional action (12) parity 
has been finally found to be 
conserved though we started with a term involving the completely 
antisymmetric tensor in 
higher dimensions. We shall comment further on this at the end.

Next, we rewrite the above equations in terms of another parameter 
$z_n~=~{\frac{m_n}{k}}e^{\sigma(\phi)}$. For the dual \KR ~field with 
masses $m_{n+}$ equation (13) takes a form which resembles the transformed 
Bessel differential equation  
\begin{equation}
\left [z_n^2\frac{d^2}{dz_n^2}+(1-\frac{2M_0}{k})z_n\frac{d}{dz_n}+ 
z_n^2\right]\chi^n=0 
\end{equation}  
\noindent
admitting a solution of  Bessel and Neuman function of order 
$\n=\frac{2M_0}{k}$ 
\begin{equation}
\chi^{n}=\frac{z_n^{\nu}}{N_n}\left[J_{\n}(z_{n})+
\alpha_{n} Y_{\n}(z_{n})\right]
\end{equation}  
\noindent
where $N_n$ is the normalisation constant and $\a_n$ are the constant 
coefficients to be determined from the boundary conditions. Likewise, in 
the anti dual case, equation (14) can also be converted into a transformed 
Bessel differential equation 
\begin{equation}
\left [z_n^2\frac{d^2}{dz_n^2}+(1+\frac{2M_0}{k})z_n\frac{d}{dz_n}+ 
z_n^2\right]\chi^n=0 
\end{equation}  
with a similar  kind of solution 
\begin{equation}
\chi^{n}=\frac{z_n^{-\nu}}{N_n}\left[J_{\n}(z_{n})+
\alpha_{n} Y_{\n}(z_{n})\right]
\end{equation}  
of order $\n=\frac{2M_0}{k}$. $N_n$ is the overall normalisation,
over and above the constant $\a_n$.  As mentioned earlier, we determine 
these constants from the boundary conditions which requires the first 
derivative of $\chi^n$ to be continuous at the orbifold fixed points 
$\phi~=~0$ and $\phi=\pm\pi$ so that the self-adjointness of the left hand 
side of equation (13) (for the dual \KR ~field or (14) in case of antidual 
field) holds. From the continuity condition at $\phi=0$ along 
with the limit $e^{kr_c\pi}>>1$ and the requirement that the  
mass values $m_n$ on the brane should be of the order of 
TeV $(<<k)$, we obtain 
\begin{equation}
\alpha_n \simeq \left[ \frac{1}{\sqrt{\n-2}{(\n-1)}!}\left(\frac{x_n}{2} 
e^{-2kr_c\pi}\right)^{\n-1}\right]^2
\end{equation} 
\noindent
for the self dual field with $x_n~=~z_n(\pi)$. Similarly,  $\a_n$ for 
the antidual field is given by
\begin{equation}
\alpha_n \simeq \left[ \frac{1}{\n!}\left(\frac{x_n}{2} e^{-2kr_c\pi}
\right)^{\n+1}\right]^2
\end{equation} 
It is quite apparent that for both the cases $\a_n<<1$ in the limit mentioned 
above. Therefore, in this approximation, the solution for $\chi^n$ 
presented in equations (16) and (18) for the two cases now becomes
\be
\chi^n=\frac{z_n^\n}{N_n}J_\n(z_n)~~~~~~{\rm and}~~~~~~~~~\chi^n=
\frac{z_n^{-\n}}{N_n}J_\n(z_n) 
\ee
respectively, where $N_n$ are the normalisation constants in the two 
cases determined by the orthonormality condition (11). Next, we consider 
the continuity condition at $\phi=\p$, which yields
\begin{equation}
J_{\n-1}(x_n)= 0~~~~~~~~{\rm and}~~~~~~~~ J_{\n+1}(x_n)= 0
\end{equation}
for the dual and anti-dual field respectively. So, $x_n$ is simply the 
root of the corresponding Bessel function. Consequently the mass spectrum 
in both the cases depends on the corresponding Bessel function. Thus, with 
values of $x_n$  of the order of unity, the mass eigenvalues of the higher 
KK excitation $\sqrt3x_nke^{-kr_c\p}$ settle at the TeV range, as expected. 
One important point to note here is that the value of $x_n$ are function 
of $\n$ i.e, the ratio of the coupling mass parameter $M_0$ to $k$. As 
mentioned in the begining both $M_0$ and $k$ are on the order of the Plank 
scale  $M_P$, but fractional difference also changes the eigenvalues of 
the mass spectrum. We present in figure 1 lowest possible massive mode 
as a function of the ratio $M_0/M_P$, with three values of the ratio
$M_P/k$ in each case. The lower limit , {\it i.e.} $(M_P/k=10)$, arises
from the constraint that the scalar curvature in 5-dimensions
exceeds $M_P^2$ for higher values of $k$.   

\begin{figure}[t]
\centering
\leavevmode \epsfysize=7cm \epsfbox{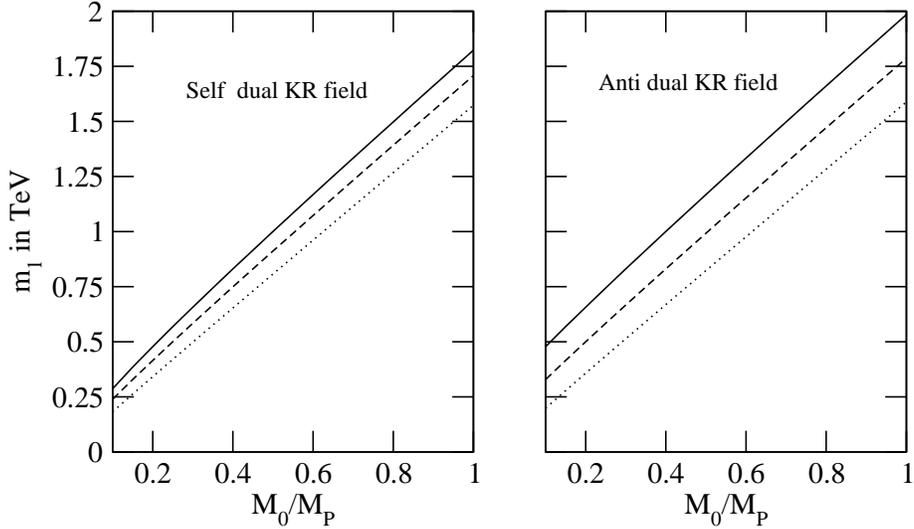}
\vskip 0.1cm
\caption{Plot of the masses $m_1$ in TeV scale with respect to 
$M_0/M_P$ where $M_P=10^{19}$ and $kr_c=12$ and
 with different ratios for $M_P/k$ (such as $M_P/k=10,20~{\rm and}~100$ 
from the upper to the lower curves respectively in both figures). The 
first figure represents the mass of the self dual field for 
the lowest root of the Bessel function $J_{\n-1}(x)=0$ where $\n=2M_0/k$, 
while the second one gives the masses of the antidual field, where 
the order of the Bessel function is $2M_0/k+1$.}  
\label{figure1}
\end{figure}

Using the expression for $\chi^n$ in (11) one can find the 
normalisation constant. For the selfdual case  
\be
N_n=\left[\int_{-\pi}^\p d\phi e^{2\sigma(\phi)} z_n^{2\n}J_\n(z_n)^2
\right]^\frac{1}{2} 
\ee
\noindent
and for the anti selfdual field the normalisation condition yields
\be
N_n=\left[\int_{-\pi}^\p d\phi e^{2\sigma(\phi)} 
z_n^{-2\n}J_\n(z_n)^2\right]^\frac{1}{2} 
\ee

So far we were confined with the massive modes of the KR field and found 
that with the inclusion of the Chern Simons like term the masses of higher 
KK modes depends on the coupling parameter $M_0$. Now,  let us find the 
solution for the massless mode of the four dimensional projection of 
the KR field for both cases. Solving equation (13) and (14)  for the 
massless KK mode, the zero mode of $\chi^n$ turn out to be 
\be
\chi^0=c_1+\frac{c_2}{2M_0r_c} e^{2M_0r_c|\phi|}~~~{\rm and}~~~\chi^0=
c_1-\frac{c_2}{2M_0r_c} e^{-2M_0r_c|\phi|}
\ee
\noindent
for the self dual and anti dual field respectively. The imposition of  
self-adjointness forces us to a constant solution for 
$\chi^0$. Using the orthonormality sondition, one thus obtains,
\be
\chi^0=\sqrt{kr_c}e^{-kr_c\p}
\ee
for both the cases. Thus the zero mode exhibits a suppression by the 
warp factor. This result is identical to the one found in \cite{prl}.  
Thus with the most general Kalb Ramond gauge invariant action also we 
still conclude that the massless KR mode is enormously suppressed on 
the visible brane, ensuring the absence of torsion.  

At this point it is quite natural to make a comparison with the 
case where the topological term is absent. As has been mentioned 
earlier, the masses now depend crucially 
on $M_0/k$ as compared to the case presented in \cite{prl} where a minimal 
torsion term was present in the 5D action along with gravity and the 
masses of the higher KK modes were given by the zeros of first order 
Bessel function $J_1(x_n)$. The inclusion of the 
topological term in the action has shifted the mass spectrum 
from the earlier case but it is still very much in the TeV range and 
within the reach of the TeV-scale collider experiments. In \cite{prl}, 
it was claimed that the  masses of the KR tower are just given by
a factor of $\sqrt3$  scaling 
from the graviton modes\cite{hrd2}, whereas in this case the coresponding 
masses also depend upon the parameter $M_0$. 
For a better comparison, in figure 2 we present some plots for the massive 
spectrum $m_n$ against $M_0 /k$ in the three cases (with the minimal 
torsion term\cite{prl}, self-dual and anti-dual KR  field). The dashed 
lines correspond to the earlier case with the minimal 
torsion term $(J_1(x_n))$, the bold lines, to the self-dual KR field 
($J_{\n-1}(x_n)$), and the dotted lines, to the anti-dual 
field ($J_{\n+1}(x_n)$, respectively). The three different plots correspond 
to the three lowest roots of the corresponding Bessel function. As is 
evident from the plot, both in the dual and the antidual cases
the mass eigenvalues are boosted with the respect to the cases in
\cite{prl}, although they are still within the kinematic reach of
TeV-scale experiments. However, their `visibility' in the scenario considered
here is drastically reduced, as discussed below.

\begin{figure}[t]
\centering
\leavevmode \epsfysize=6.5cm\epsfbox{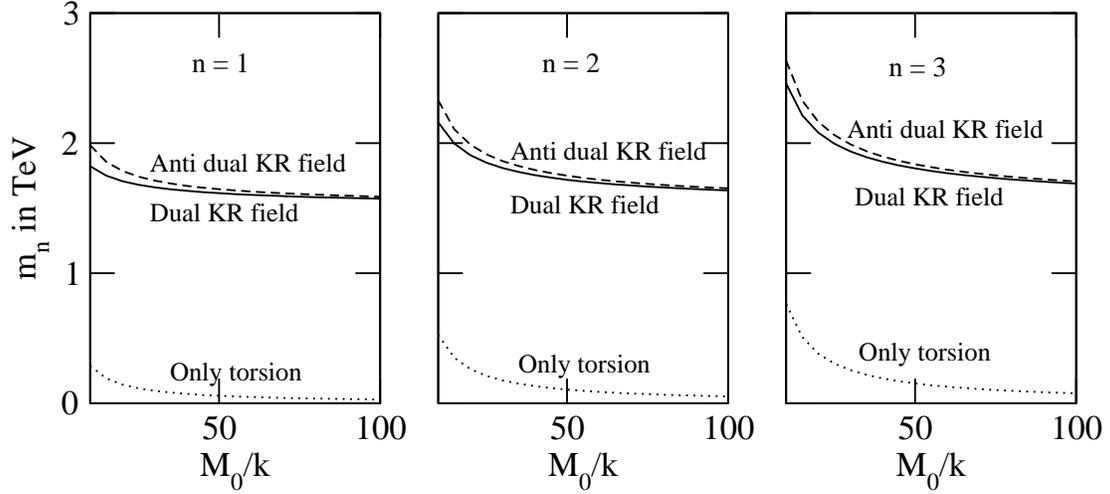}
\caption{Plot of the three lowest mass eigenvalues of the 
higher KK modes with respect 
to $M_0/k$ for the three lowest root of the Bessel function. The lowest line 
(dotted line) correspond to the case with torsion only\cite{prl} 
with $J_1(x)=0$. The second line from the bottom (bold line) correspond 
to the case with dual KR field ($i.e, J_{\n-1}(x)=0$ where $\n=2M_0/k$) 
and the uppermost line (dashed line) correspond to the antidual KR field 
($i.e, J_{\n+1}(x)=0$). Here also  we have chosen $M_0=M_P=10^{19}$, 
with $kr_c~=~12$.}
\label{figure3}
\end{figure}

As has already been indicated, the couplings of the 
massless modes with matter  modes in this case are of the same kind as
in reference \cite{prl}. For massive fields, the difference caused by the
introduction of the $\epsilon BH$ term becomes obvious if we  consider the
interactions with fermion fields, and compare them with equation (26)
of \cite{prl}. 

For self-dual KR field, the coupling looks as follows:

\begin{equation}
{\cal L}_{\psi\bar{\psi}H}= - \bar{\psi}[i\gamma^{\mu}\sigma^{\nu\lambda}
\left\{\frac{1}{M_P
e^{kr_c\pi}}H^0_{\mu\nu\lambda}+\frac{1}{\sqrt{12}M_P}
\sum_{n=1}^{\infty}\frac{x_n^\nu}{N_n}
J_\nu(x_n)H^n_{\mu\nu\lambda}\right\}]\psi
\end{equation}

\noindent
while for an antidual field, the expression is

\begin{equation}
{\cal L}_{\psi\bar{\psi}H}= - \bar{\psi}[i\gamma^{\mu}\sigma^{\nu\lambda}
\left\{\frac{1}{M_P
e^{kr_c\pi}}H^0_{\mu\nu\lambda}+\frac{1}{\sqrt{12}M_P}
\sum_{n=1}^{\infty}\frac{x_n^{-\nu}}{N_n}
J_\nu(x_n)H^n_{\mu\nu\lambda}\right\}]\psi
\end{equation}

\noindent
using $kr_c~=~12$. 

The remarkable point to note here is that the massive mode coupling to
fermion, as given by equations (27) and (28), are drastically reduced 
compared to the corresponding case {\it without the presence of the
$\epsilon BH$ term}. Basically, the different co-efficients
arising from the solutions of equations (13) and (14), are responsible for
this. This feature has its origin in the second  term on the
left-hand side of each of them, which in turn owes itself to the last term
in the action given in equation (5). It appears as if the large
coefficient $M_0$ in the additional five dimensional topological term 
$\epsilon BH$  causes
the KR modes to decouple from all visible physics on the brane, although
a tower within the kinematic reach of accelerator experiments 
is still around.

The analysis presented can be summarised as follows: We have considered 
the most general five-dimensional action invariant under Kalb Ramond gauge 
transformation where gravity and torsion are accorded the same status in 
the bulk. This is a kind of the generalisation of the action presented in 
\cite{prl}. The compactification of the fifth dimension gives rise to a 
spectrum of Kaluza Klein modes for the KR field. On solving the equation 
of motion in four dimensions, we find that the only non-trivial classical 
solutions correspond to self-dual or anti-dual KR field. This 
result turns out to be 
crucial for the effective four dimensional theory in the following ways:

(a)In spite of the inclusion of a five dimensional Levi-Civita 
term in the action, the effective four dimensional action after 
Randall-Sundrum compactification continues to be  parity invariant.
One expects that a parity violating interaction will be generated from 
the 4-dimensional Levi-Civita dependent term which appears from the 
five-dimensional Levi-Civita term after the RS compactification.
However the KR field equations constrains the KR field to be selfdual or 
anti-selfdual only and thereby reduces the number of degrees of freedom
and thus removes the compactified four dimensional Levi-Civita from the 
effective 4d action.
As a result the effective 4d theory continues to be parity invariant.
This is in perfect agreement with our earlier findings that parity seems 
to be protected by duality in all such scenarios\cite{bs,bsss,prd}. (b)The 
massive KK modes spectrum for both cases depends on the coupling parameter 
$M_0$ and are shifted from the earlier case as expected, due to the inclusion 
of the additional term. But the mass eigenvalues of the spectrum are still 
on the order of TeV. However, the massive modes now have drastically
reduced interaction strength with matter fields,  and this 
practically destroys the scope for their detection in
accelerator experiments. Such a reduction of coupling of the massive 
modes stems from selfdual or anti-selfdual nature of the KR field which 
has originate from the additional topological term with the large 
coefficient $M_0$ in the action. In other words, whereas a minimal
theory \cite{prl} predicts an invisible zero-mode torsion but
envisions potentially observable massive modes, the existence of
the $\epsilon BH$-term makes all modes, massless or massive, 
practically invisible. 
However, the most significant result obtained in \cite{prl}, namely, 
the massless mode of the four dimensional projection 
of the Kalb Ramond field  
has an added suppression through the warp factor with respect to gravity 
in its interaction with other matter field remains valid in 
this modified scenario also. 
In addition the present scenario restricts the 
KR field to be selfdual or anti-selfdual. Therefore, starting from 
this more generalised action in five dimensions we still have a 
parity-conserving effective  torsionless universe when 
we consider the projection of massless torsion mode on 
the visible brane.

\noindent
{\bf Acknowledgement:} The work of BM has been partially supported by
the Board of Research in Nuclear Sciences, Government of India. The work of 
SS is financed by FCT, Portugal, through CAAUL.

\noindent
{\bf Appendix:}

\noindent
In the appendix we derive the solution of the KR field on solving the 
equation of motion in 4D. After integrating out the extra dimension we 
are left with the action (10) in the effective four dimension. The 
Lagrangian formed from action (10) is the following,
\be
{\cal{L}}_B=\partial_{[\m}B^n_{\n\l]}\partial^{[\m}B^{n~\n\l]}-
\s^2~B^n_{\m\n}B^{n~\m\n}+ C~B^n_{\m\n}{\tilde B}^{n~\m\n}
\ee
\noindent
where $\tilde{B}^{n~\mu\nu}~(=\epsilon^{\mu\nu\alpha\beta}B^n_{\alpha\beta})$
 is the dual of \KR ~field and $\s$ and $C$ are two constants defined by
\noindent
$$
\frac{e^{-2\sigma}}{r_c^2}~\frac{1}{\chi^n}\partial_\phi\partial_\phi\chi^n=
\s^2 $$
\noindent
and 
$$
\frac{e^{-2\sigma}}{r_c}~\frac{M_0}{\chi^n}\partial_\phi\chi^n=C. $$

The equation of motion obtained from the KR action is
\be
(\Box-\s^2)B^{n~\m\n}=C{\tilde B}^{n~\m\n}
\ee
\noindent 
From now onwards we use the notation $\star$ to denote the duality 
operation such that $\star B_{\m\n}={\tilde B}_{\m\n}$. Then the above 
equation can be written as,
\be
(\Box-\s^2)B^{n~\m\n}=C~\star B^{n~\m\n}
\ee

\noindent
Next we make use of a well known result of differential geometry. 
Namely, on any manifold 
\be
\Box~\star~=~\star~\Box
\ee
\noindent
That is to say, the duality operation commutes with the d'Alambertian. 
Now, applying $(\Box-\s^2)$ once more on equation (31) and using the above 
result, we have,
\bea
(\Box-\s^2)^2 B^{n~\m\n}&=&C(\Box-\s^2)~\star B^{n~\m\n}\nonumber\\
&=&C~\star(\Box-\s^2)B^{n~\m\n}\nonumber\\
&=& C^2 B^{n~\m\n}
\eea

\noindent  
which means,
\be
(\Box-\s^2+C)(\Box-\s^2-C)B^{n~\m\n}=0
\ee

\noindent
Keeping in mind that the two factors on the left hand side of the above 
equation commutes, several possibility emerges out of the above equation. 
We take them up one by one.

{\em Possibility I}: \be
(\Box-\s^2-C)B^{n~\m\n}=0
\ee
\noindent 
which implies 
\be
B^{n~\m\n}=\star B^{n~\m\n}~~~{\rm i.e,}~~B^{n~\m\n}~~{\rm is~~self dual.}
\ee

{\em Possibility II}:
\be
(\Box-\s^2+C)B^{n~\m\n}=0
\ee
\noindent
implies
\be
-B^{n~\m\n}=\star B^{n~\m\n}~~~{\rm i.e,}~~B^{n~\m\n}~~{\rm is~~anti~~
self dual.}
\ee

{\em Possibility III}: The third possibility arises when none of the factors 
in equation (34) operating on $B^{n~\m\n}$ gives zero, yet the product is 
zero. This could happen with operators. In such case both of the following 
conclusion will hold.

(a) Defining
\be
(\Box-\s^2+C)B^{n~\m\n}=F
\ee
\noindent 
one gets
\bea 
C~\star B^{n~\m\n}+C~B^{n~\m\n}=F\nonumber\\
F=\star~F
\eea
\noindent 
Now we take the dual of equation (39)
\be 
(\Box-\s^2)\star B^{n~\m\n}+ C~\star B^{n~\m\n}=\star F
\ee
\noindent
But equation (40) and (41) together mean
\bea
(\Box-\s^2+C)B^{n~\m\n}=(\Box-\s^2+C)~\star B^{n~\m\n}\nonumber\\
{\rm ~~i.e,~~~~~~~~}B^{n~\m\n}~=~\star B^{n~\m\n}
\eea

(b) At the same time we define 
\be
(\Box-\s^2-C)B^{n~\m\n}=F^{\prime}
\ee
Similarly one gets
\bea 
C~\star B^{n~\m\n}-C~B^{n~\m\n}=F^{\prime}\nonumber\\
\star~F^{\prime}=-~F^{\prime}
\eea
\noindent
Again taking the dual of equation (43)
\be 
(\Box-\s^2)\star B^{n~\m\n}- C~\star B^{n~\m\n}=\star F^{\prime}
\ee
\noindent
which together with (44) implies
\bea
-~(\Box-\s^2-C)B^{n~\m\n}=(\Box-\s^2-C)~\star B^{n~\m\n}\nonumber\\
{\rm ~~i.e,~~~~~~~~}-B^{n~\m\n}~=~\star B^{n~\m\n}
\eea

Now from (a) and (b) above, if both the equation (42) and (46) has to 
follow the only conclusion is that $B^{n~\m\n}=0$. Note that this will 
not be the case if either $F$ or $F^{\prime}$ is zero, in which case we 
get back the possibility I or II. So, combining all the possibilities, we 
could conclude that with the term $C~B^n_{\m\n}{\tilde B}^{n~\m\n}$ present 
in the 4 dimensional lagrangian we must have,
\bea
{\tilde B}^{n~\m\n}=B^{n~\m\n}\nonumber\\
{\tilde B}^{n~\m\n}=-~B^{n~\m\n}\nonumber\\
 B^{n~\m\n}~=~0
\eea
\noindent 
That is to say the only nontrivial solutions correspond to selfdual or 
anti-selfdual \KR ~fields.

\end{document}